\documentclass[twocolumn,showpacs,preprintnumbers,amsmath,amssymb]{revtex4}

\usepackage{graphicx}
\usepackage{dcolumn}
\usepackage{bm}

\begin{document}


\title{
Emergence of fractal in aggregation with stochastic self-replication
}

\author{Md. Kamrul Hassan$^1$, Md. Zahedul Hassan$^2$ and Nabila Islam$^1$}

\affiliation{
$1$ Department of Physics, University of Dhaka, Dhaka 1000, Bangladesh \\
$2$ Institute of Computer Science, Bangladesh Atomic Energy Commission, Dhaka 1000, Bangladesh
}

\begin{abstract}

We propose and investigate a simple model which describes the kinetics of aggregation of Brownian particles with stochastic 
self-replication. An exact solution and the scaling theory are presented alongside numerical simulation which fully 
support all theoretical findings. In particular, we show analytically that the particle size distribution
function exhibits dynamic scaling and we verified it numerically using the idea of data-collapse. Besides, the 
conditions under which the resulting system emerges as a fractal are found, the fractal dimension of the system is 
given and the relationship between this fractal dimension and a conserved quantity is pointed out.
\end{abstract}



\maketitle

\section{Introduction}
The kinetics of irreversible aggregation of particles is one of the most fundamental yet 
challenging and fascinating problems. It occurs in a variety of processes in physics, chemistry, biology and engineering.
For instance, aggregation of colloidal or aerosol particles suspended in liquid or gas, 
polymerization, antigen-antibody aggregation and cluster formation in galaxy etc. are just a few examples
to name \cite{ref.friedlander,ref.johnstone,ref.silk}.
A comprehensive description
of the aggregation process which takes into account the sizes or
masses, positions, velocities, geometries and reaction mechanisms of
the aggregating particles is a formidable problem and presently beyond the scope of precise
theoretical analysis. The best that can be achieved analytically till to date is to
characterize aggregating particles according to their sizes
or masses only and describe the process via a kinetic reaction scheme, 
\begin{equation}
\label{eq:reaction}
 A_{x}(t) + A_{y}(t)\stackrel{R} {\longrightarrow} A_{(x+ y)}(t + \tau).
\end{equation}
Here, $A_{x}(t)$ represents an aggregate of size $x$ at
time $t$ and $R$ is the rate at which aggregates of size $x$ at time $t$
joins irreversibly  with another particle of size
$y$ upon encounter and form a new particle of size $(x+y)$. 

The time-evolution of a system of chemically identical
particles which obey the reaction scheme given by Eq. (\ref{eq:reaction}) can be well described by
Smoluchowski's equation \cite{ref.smoluchowskiSol,ref.chandrasekhar}
\begin{eqnarray}
\label{eq:smoluchowski}
& & {{\partial c(x,t)}\over{\partial t}}= -c(x,t)\int_0^\infty K(x,y)c(y,t)dy 
\nonumber \\ & + &{{1}\over{2}} \int_0^x K(y,x-y) c(y,t)c(x-y,t)dy.
\end{eqnarray}
In this equation, $c(x,t)$ is the concentration of particles of size
$x$ at time $t$ and $K(x,y)$ is the kernel that determines the rate at which particles of size $x$
and $y$ combine to form a particle of size $(x+y)$ since the reaction rate is given by $R={\int_0^\infty
K(x,y)c(y,t)dy}$. On the other hand, the factor $1/2$ in the gain term implies that at each step two particles 
combine to form one particle. The Smoluchowski equation has been studied extensively in and around the eighties for a 
large class of kernels satisfying  $K(bx,by)=b^\lambda K(x,y)$,  
where $b>0$ and $\lambda$ is the homogeneity index. Significant contributions towards the understanding of the scaling
theory and sol-gel phase transitions have been made during this period
\cite{ref.ziff,ref.ziffEtAl,ref.vanDongenErnst,ref.leyvraz}.

Much of the recent theoretical work on aggregation has been devoted to making the Smoluchowski
equation more versatile. This is mainly driven by the thirst of gaining deeper insight into the 
systems beyond the scope of the Smoluchowski equation. For instance, Krapivsky and Ben-Naim proposed a
model that involves aggregation of two types of particles, active and passive, in an attempt to 
explain multi-phase coarsening processes and polymerization of linear polymers 
\cite{ref.krapivsky, ref.ben-naim}.  Ke {\it et al.} proposed yet another aggregation model
with monomer replications and/or self-replications intended to
explain processes such as DNA replication \cite{ref.keEtAl}.  
Besides, Hassan and Hassan recently proposed a model that considers aggregation of particles growing by
heterogeneous condensation and shown that the resulting system emerges as fractal which is accompanied by the 
violation of conservation of mass \cite{ref.CDA, ref.fractalCDA}. To the best of our knowledge this has been the only analytcal work that found 
fractal in aggregation process albeit there exist numerous laboratory experiments and numerical simulations which suggested that scale-invariant
fractals almost always emerge when particles aggregate \cite{ref.botet, ref.vicsek}. We need more exactly
solvable analytical models to elucidate and
explain why fractals are so ubiquitous in aggregation processes. 
The present work therefore can be seen as yet another attempt to that end. 

In this work, we propose a very simple variant of
the Smoluchowski equation in which we investigate aggregation of
particles accompanied by self-replication of the newly formed
particles with a given probability $p$. The spirit of our model, in some senses, is similar to that of 
the work of Ke {\it et al.} \cite{ref.keEtAl}.
In contrast to their work where self-replication is facilitated  
by a rate kernel, in our case self-replication is facilitated by a prior choice
of the probability $p$. Besides, we may consider that the system of our model has two different kinds of 
particles: active and passive.
As the systems evolves, active particles always
remain active and take part in aggregation while the character of the passive particles are altered irreversibly
to an active particle with probability $p$. Once a passive particle turns into an active particle it 
can take part in further aggregation like other active particles already
present in the system on an equal footing and never turns into a passive particle. 
This interpretation is very similar to the work of Krapivsky and Ben-Naim \cite{ref.krapivsky, ref.ben-naim}.
While in their work the character of an active particle is altered, in our work it is the other way around.  
The two models are different also because here we only consider the dynamics of the active particles,
whereas Krapivsky and Ben-Naim studied the dynamics of both the entities since a passive
particle in their case exist at the expense of an active particle and therefore a consistency check is required. However, 
the present model does not require such consistency check. 

There are many real physical
systems where both aggregation and self-replication  occur naturally. For instance, the
symbiosis-driven growth of biological systems, the
replication-driven amplification of cells and DNA replication in
polymerase chain reactions \cite{ref.cairns, ref.lipps,
ref.stolovitzkyCecchi}. The model we propose can also describe systems where passive clusters coexist with active cluster
without disturbing the dynamics of the latter. For instance, in polymerization of linear polymers the system may contain
chemically active as well as initially inert (or passive) polymers of poly-disperse distribution of sizes.
Active and passive clusters can also co-exist in multiphase coarsening processes in one dimension whereby upon
merging the domain walls may remain active or become passive depending on the surface tension of the phase
of the neighboring domains. Besides its potential application in various physical processes, 
it is also interesting from the pedagogical point of view as it is an exactly solvable 
analytical model that can interpolate between stochastic fractal with tunable fractal
dimensions for $0<p<1$ and Euclidean dimensions for $p=0$.

The rest of the paper is organized as follows. In section II, we present the definition of our model and the generalized Smoluchowski equation
that can describe the model. In section III, we give an exact solution to the generalized Smoluchowski equation valid for all time $t$. 
The scaling theory of the Smoluchowski equation we propose is discussed in section IV. In section V, we invoke the idea of fractal analysis to 
give a geometric interpretation of our model. Finally, in section VI we give a general discussion and summary of the work.

\section{The model}

 Perhaps an exact algorithm can provide a better description of the model than its mere definition.
The process starts with a system that comprise of a large number of chemically identical Brownian particles  
and a fixed value for the probability $p \in [0,1]$ by which particles are self-replicated. 
The alogorithm of the model can then be described as follows: 
\begin{itemize}
\item[(i)] Two particles, say of sizes $x$ and $y$, are picked randomly from the system to mimic a random collision 
via Brownian motion.
\item[(ii)] Add the sizes of the two particles to form one particle of 
their combined size $(x+y)$ to mimic aggregation. 
\item[(iii)] Pick a random number $0<R<1$. If $R\leq p$ then add another particle of size $(x+y)$ to the system to
mimic self-replication.  
\item[(iv)] The steps (i)-(iii) 
are repeated {\it ad infinitum} to mimic the time evolution. 
\end{itemize}
Note that random collision due to Brownian motion can be ensured if we choose a constant kernel $K(x,y)$, e.g.
\begin{equation}
\label{eq:kernel}
    K(x,y) = 2,
\end{equation}
for convenience. The Smoluchowski equation with constant kernel then corresponds to the $p=0$ case.   
On the other hand, the other extreme $p=1$ case describes the fact that whenever two particles, say of size $x$ and $y$,
come into contact they form a particle of their combined size $(x+y)$ 
and at the same time a particle of size $(x+y)$ is replicated. That is, in this case two particles always
becomes two and hence the factor $1/2$ in the gain term of the Smoluchowski equation
has to be replaced by a factor of $2/2=1$. We now consider the case where 
this latter process occurs with some probability
$p \in [0,1]$ and aggregation without replication occurs with probability $(1-p)$. Combining the two processes
we can immediately write the following  generalized Smoluchowski equation  
\begin{eqnarray}
\label{eq:modifiedSmoluchowski}
    {\frac{\partial c(x,t)}{\partial t}} &=& - 2c(x,t)\int_0^\infty dy c(y,t)
 + (1+p)\nonumber \\ &\times& \int_0^x dy c(y,t)c(x-y,t).
\end{eqnarray}
This is the fitting equation to the model described by the algorithm $(i)-(iv)$ and
the reaction scheme 
\begin{equation}
\label{eq:stochasticReaction}
    A_{x}(t) + A_{y}(t)\stackrel{R}{\longrightarrow}(1+p)A_{(x+y)}(t+\tau).
\end{equation}

\section{An exact solution}

To gain some insights of the problem we first define 
the $j \rm {th}$ moment
$M_j(t)$ of $c(x,t)$ by
\begin{equation}
\label{eq:moment}
    M_j(t)=\int_0^\infty x^{j}c(x,t)dx
\end{equation}
where $j$ is real and $j\geq 0$. Differentiating $M_j(t)$
with respect to $t$ and using Eq.
(\ref{eq:modifiedSmoluchowski}) we obtain
\begin{eqnarray}
\label{eq:dMomentdt}
{{dM_j(t)}\over{dt}} & = & \int_0^\infty\int_0^\infty dx dy c(x,t)c(y,t) \\ \nonumber
& \times & \Big [(1+p)(x+y)^j-x^j-y^j)\Big].
\end{eqnarray}
Setting $p=0$ and $j=1$ we can recover the conservation of mass ($M_1(t)={\rm const.}$) of the classical 
Smoluchowski equation for constant kernel. It is clearly evident from Eq. (\ref{eq:dMomentdt}) 
 that the mass of the system for $0<p<1$ is no longer a conserved 
quantity, and it is obvious due to the inherent definition of our model. However, it is not obvious 
from Eq. (\ref{eq:dMomentdt}) if the system is still governed by the conservation 
law or not.
Note that Eq. (\ref{eq:modifiedSmoluchowski}) essentially describes the Brownian aggregation since particles
follow Brownian motion with constant diffusivity regardless of the size. 
Whenever two such Brownian particles come into contact they merge
irreversibly to form a particle of their combined size and at the same time a particle of the same size 
is replicated with prbability $p$ revealing that the conservation of mass principle is violated.

The solutions to Eq. (\ref{eq:dMomentdt}) for the first two moments, 
namely $M_0(t)\equiv N(t)$ and $M_1(t)\equiv L(t)$, for the mono-disperse
initial condition $c(x,0)=\delta(x-1)$ are
\begin{equation}
\label{eq:m0}
N(t)= \frac {1}{(1+(1-p)t)},
\end{equation}
and
\begin{equation}
\label{eq:m1}
    L(t) = L(0){(1 + (1 - p)t)}^{\frac{2p}{1-p}}, \hspace{0.45 cm} 0\leq p < 1,
\end{equation}
respectively. To solve Eq. (\ref{eq:modifiedSmoluchowski}) we now use a 
Laplace transform $\phi(k,t)$ of $c(x,t)$ with respect to $x$ and find that
 $\phi(k,t)$ satisfies
\begin{equation}
\label{eq:dphidt}
    {\frac{\partial \phi(k,t)}{\partial t}} =-2N(t)\phi(k,t)
    +((1 + p)\phi^2(k,t),
\end{equation}
where $N(t)$ is given by Eq. (\ref{eq:m0}).
To solve Eq. (\ref{eq:dphidt}) exactly we use mono-disperse initial condition
\begin{equation}
\label{eq:initialConditionsPhi}
    \phi(k,0) = \int_0^\infty dx e^{-kx} \delta(x-1)=e^{-k},
\end{equation}
and linearize Eq. (\ref{eq:dphidt}) by making a transformation of
the form $\phi(k,t) = 1/\psi(k,t)$ to obtain
\begin{equation}
 {\frac{\partial \psi(k,t)}{\partial t}} -{{2}\over{1+(1-p)t}}\psi(k,t)=-(1+p).
\end{equation}
Then using the idea of integrating factor method we can re-write it as
\begin{equation}
 {\frac{\partial}{\partial t}}\Big ({{\psi(k,t)}\over{(1+(1-p)t)^{2/(1-p)}}}\Big )=-{{1+p}\over{(1+(1-p)t)^{2/(1-p)}}}.
\end{equation}
Integrating and then going back to the original function $\phi(k,t)$ we obtain the following solution for the
mono-disperse initial condition
\begin{equation}
\label{eq:phi}
    \phi(k,t) ={{1}\over{(1+(1-p)t)^{2/q}\Big[e^k-\Big ( 1- {{1}\over{(1+(1-p)t)^{{{1+p}\over{(1-p)}}}}}\Big )\Big ]}}.
\end{equation}
Using it in the definition of the inverse Laplace
transform  and a subsequent short calculation yields
\begin{equation}
\label{eq:c}
c(x,t) = {{1}\over{(1+(1-p)t)^{2/q}}}\Big ( 1-{{1}\over{(1+(1-p)t)^{{{1+p}\over{(1-p)}}}}}\Big )^{x-1}.
\end{equation}
It may be noted that in the limit $p \rightarrow 0$, we can still
recover the solution of Smoluchowski equation \cite{ref.smoluchowskiSol}.

\begin{figure}
\includegraphics[width=8.5cm,height=5.0cm,clip=true]{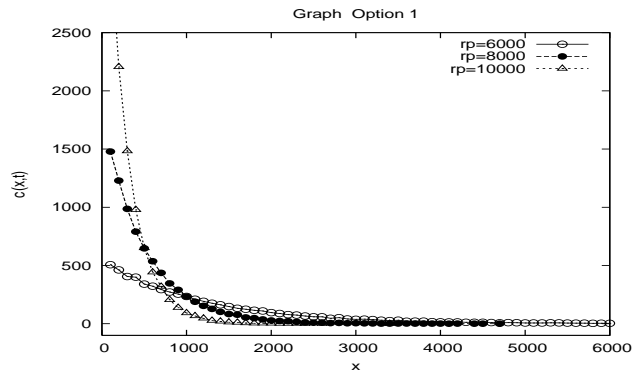}
\caption{Plot of distribution function $c(x,t)$ as a function of $x$ is shown at three different times using data obtained by numerical simulation. 
Essentially, it is a plot of a histogram where the
number of particles in each class size is normalized by the width $\Delta x$ of the interval size.
} 
\label{Figure 1}
\end{figure}

\begin{figure}
\includegraphics[width=8.5cm,height=5.0cm,clip=true]{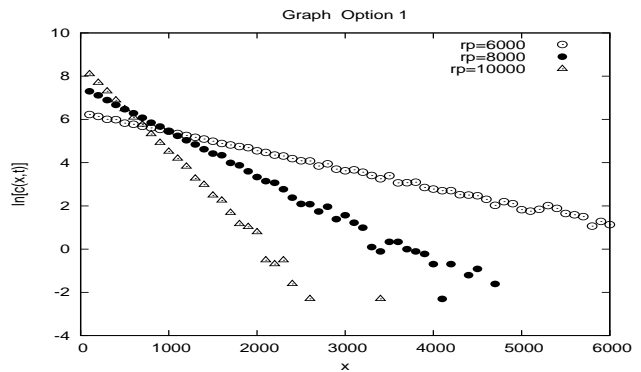}
\caption{Log-linear plot of the same data as in Fig. (1) showing the exponential decay of the particle size distribution function 
 $c_t(x)$  with particle size $x$ at fixed time as seen analytically.
}
\label{Figure 2}
\end{figure}

Of considerable interest is the long-time ($t \rightarrow
\infty$) and large-size ($x\longrightarrow \infty$) limit where the distribution function self-organize to a simpler
form. Using the long-time and large-size limit as well as the identity
\begin{equation}
{{1}\over{e}}=\lim_{n\longrightarrow \infty} \Big ( 1-{{1}\over{n}}\Big )^n,
\end{equation}
we can immediately show that the solution indeed assumes a simpler form 
\begin{equation}
\label{eq:exact_scaling_solution}
c(x,t)\sim ((1-p)t)^{-{{2}\over{(1-p)}}}e^{ -x/((1-p)t)^{ {{1+p}\over{(1-p)}}}}.
\end{equation}
This solution, however, is obtained for the mono-disperse initial condition. Consider that we have a system that
contain initially $N_0$ ($N_0\rightarrow \infty$)  chemically identical particles and 
allow them to evolve following the rules depicted in the algorithm (i)-(iv). As the process continues, we collect data at
three different instant, say at $t_1$ ,$t_2$ and $t_3$ such that $t_1<t_2<t_3$, and plot a histogram
where the number of particles in each class is normalized by the width $\Delta x$ of the interval size.
The resulting curves shown in Fig. (1) represent distribution function $c(x,t)$ vs $x$ 
at three different times $t_1$ ,$t_2$ and $t_3$. Note that each curve actually distribution function 
$c_t(x)\sim e^{ -x/((1-p)t)^{ {{1+p}\over{(1-p)}}}}$ at 
a fixed time $t$ and hence plots of $\log[c_t(x)]$ versus
$x$ should result in a straigtline with decreasing slopes (see Fig. 2). 

\section{Scaling theory}

We find it convenient first to find how the mean or typical particle size $s(t)$ grows with time $t$ as a result of
random sequential aggregation with self-replication. This is
defined as
\begin{equation}
\label{eq:meanSizeDef}
    s(t) = <x> = \frac{\int_0^\infty dx xc(x,t)}{\int_0^\infty dx c(x,t)} = \frac{M_{1}(t)}{M_{0}(t)}.
\end{equation}
Using Eqs. (\ref{eq:m0}) and (\ref{eq:m1}) we find
\begin{equation}
\label{eq:meanSize}
    s(t) =\frac{L(0)}{N(0)}{(1 + (1 - p)N(0)t)}^{(\frac{1+p}{1-p})}, \hspace{0.45 cm} 0\leq p < 1.
 \end{equation}
 We thus see that for $0\leq p<1$  
the mean particle size $s(t)$ in the limit $t \rightarrow \infty$ grows following
power-law 
\begin{equation}
\label{eq:meanSizeGrowth}
    s(t) \sim ((1-p)t)^{\frac{1+p}{(1-p)}}.
\end{equation}
To verify this we plot $\ln(s(t))$ against
$\ln(t)$ in Fig. (1) for three different values of $p$ with the same
mono-disperse initial condition in each case. Appreciating the fact that $t\sim 1/N$ in the long-time limit
we obtain three straight lines whose gradients are given by $(\frac{1+p}{(1-p)})$, providing
numerical confirmation of the theoretically derived result given by
Eq. (\ref{eq:meanSizeGrowth}).

\begin{figure}
\includegraphics[width=8.5cm,height=5.0cm,clip=true]{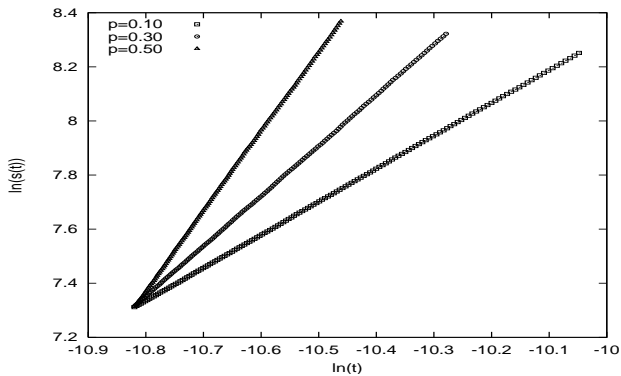}
\caption{ We plot $\ln(s(t))$ against $\ln(t)$ for three different
values of $p$ starting with mono-disperse initial conditions (we choose $50,000$ particles of unit size). The
lines have slopes given by the relation $\frac{1+p}{1-p}$,
confirming that $s(t) \sim 
t^{\frac{1+p}{1-p}}$.} 
\label{Figure 3}
\end{figure}

We shall now apply the Buckingham Pi theorem to obtain scaling solution as it will provide deeper insight into the problem \cite{ref.barenblatt}.
Note that according to Eq. (\ref{eq:modifiedSmoluchowski}) the governed parameter $c$ depends on three parameters
$x$, $t$ and $p$. However,
the knowledge about the growth law for the mean particle size implies that one of the parameters, say $x$, can be expressed
in terms of $t$ and $p$ since according to Eq. (\ref{eq:meanSizeGrowth}) the quantity $((1-p)t)^{\frac{1+p}{(1-p)}}$ bear the dimension of 
particle size. Note though that $p$ itslf does not have dimension, yet we are keeping it as we find it convenient 
for our future discussion. If we consider $(1-p)t$ as an independent parameter then the distribution function $c(x,t)$ 
too can be expressed
in terms of $(1-p)t$ alone, and using the power-law monomial nature of the dimension of physical quantity we can write 
$c(x,t)\sim ((1-p)t)^\theta$. We therefore can define a dimensionless governing parameter
\begin{equation}
\xi ={{x}\over{((1-p)t)^z}},
\end{equation}
where $z={{1+p}\over{(1-p)}}$ and a dimensless governed parameter
\begin{equation}
\label{eq:Pi_1}
\Pi={{c(x,t)}\over{((1-p)t)^\theta}}.
\end{equation}
The numerical value of the right hand side of the above two equations remain the same even if the time $t$ is changed by
some factor $\mu$ for example since the left hand side are dimensionless. It
means that the two parameters $x$ and $t$ must combine to form a dimensionless quantity 
$\xi=x/t^z$ such that the dimensionless governed parameter $\Pi$ can only depends on $\xi$. In other words,
we can write 
\begin{equation}
\label{eq:Pi_2}
{{c(x,t)}\over{((1-p)t)^\theta}}=f(x/t^z),
\end{equation}
which lead to the following dynamic scaling form
\begin{equation}
\label{eq:scaling_form}
c(x,t)\sim ((1-p)t)^\theta f(x/((1-p)t)^z), 
\end{equation}
where the exponents $\theta$ and $z$ are fixed by the dimensional relations $[t^\theta]=[c]$ and $[t^z]=[x]$ 
respectively and $f(\xi)$ is known as the scaling function \cite{ref.family_Vicsek}.

We now use the scaling form given by Eq. (\ref{eq:scaling_form}) into Eq.
(\ref{eq:modifiedSmoluchowski}) and find that the scaling function $\phi(\xi)$ satisfies
\begin{eqnarray}
\label{eq:scaling}
t^{-(\theta+z+1)} & = & {{(1-p)^{2\theta +z}}\over{F(p,\xi)}}{\Big [} -2 \mu_0f(\xi) 
\nonumber \\ & +& (1+p)\int_0^\xi f(\eta)f(\xi-\eta)d\eta {\Big ]},
\end{eqnarray}
where
\begin{equation}
F(p,\xi)=\Big[ \theta (1-p)^\theta f(\xi)-z(1-p)^\theta \xi{\frac{df(\xi)}{d\xi}}\Big ],
\end{equation}
and 
\begin{equation}
\mu_{0} = \int_{0}^{\infty}d\xi f(\xi),
\end{equation}
is the zeroth moment of the scaling function. The right hand side of Eq.
(\ref{eq:scaling}) is dimensionless and hence dimensional consistency requires $\theta+z+1=0$ or
\begin{equation}
\label{eq:theta}
    \theta =- \frac{2}{1-p}.
\end{equation}
The equation for the scaling function $f(\xi)$ which we have to solve for this $\theta$ value is
\begin{equation}
(1+p)\Big [\xi{{df(\xi)}\over{d\xi}}+\int_0^\xi f(\eta)f(\xi-\eta)d\eta\Big ] =
2f(\xi)(\mu_0-1).
\end{equation}
Integrating it over $\xi$ from $0$ to $\infty$ immediately gives $\mu_0=1$ and hence
the equation that we have to solve to find the scaling function $f(x)$ is
\begin{equation}
\label{eq:scaling_phi}
\xi{{df(\xi)}\over{d\xi}}=-\int_0^\xi f(\eta)f(\xi-\eta)d\eta.
\end{equation}

To solve Eq. (\ref{eq:scaling_phi}) we apply the Laplace transform $G(k)$ of $f(\xi)$ in Eq. (\ref{eq:scaling_phi}) and find that $G(k)$
satisfies 
\begin{equation}
{{d}\over{dk}}\Big ( kG(k)\Big )=G^2(k).
\end{equation}
It can be easily solved after linearizing it by making transformation of the form $G(k)=1/u(k)$ and integrating
straighaway gives 
\begin{equation}
G(k)={{1}\over{1+k}}.
\end{equation}
Using it in the definition of the inverse Laplace transform we find the required solution
\begin{equation}
\label{eq:scaling_function}
f(\xi)=e^{-\xi},
\end{equation}
and hence accordng to Eq. (\ref{eq:scaling_form}) the scaling solution for the distribution function is
\begin{equation}
\label{eq:scaling_solution}
c(x,t)\sim ((1-p)t)^{-{{2}\over{(1-p)}}}e^{ -x/((1-p)t)^{ {{1+p}\over{(1-p)}}}}.
\end{equation}
It is exactly the same as in Eq. (\ref{eq:exact_scaling_solution}).  The advantage of using the scaling theory 
is that one does not need to specify the initial condition revealing the fact that the solution is true
for any initial condition.

\begin{figure}
\includegraphics[width=8.5cm,height=5.0cm,clip=true]{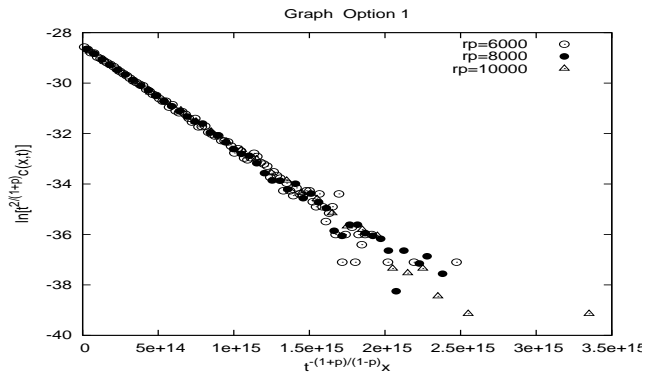}
\caption{The three distincts curves of Figs. (1) and (2) for three different system sizes are well 
collapsed onto a single universal curve 
when $c(x,t)$ is measured 
in units of $t^{-{{2}\over{(1-p)}}}$ and  $x$ measured in units of $t^{{{1+p}\over{(1-p)}}}$. Such data-collapse 
implies that the process evolves with time preserving its self-similar character.
We have chosen semi-log scale to demonstrate
that the scaling function decays exponentially $f(\xi)\sim e^{-\xi}$ as predicted by the theory. 
} 
\label{Figure 4}
\end{figure}

\begin{figure}
\includegraphics[width=8.5cm,height=5.0cm,clip=true]{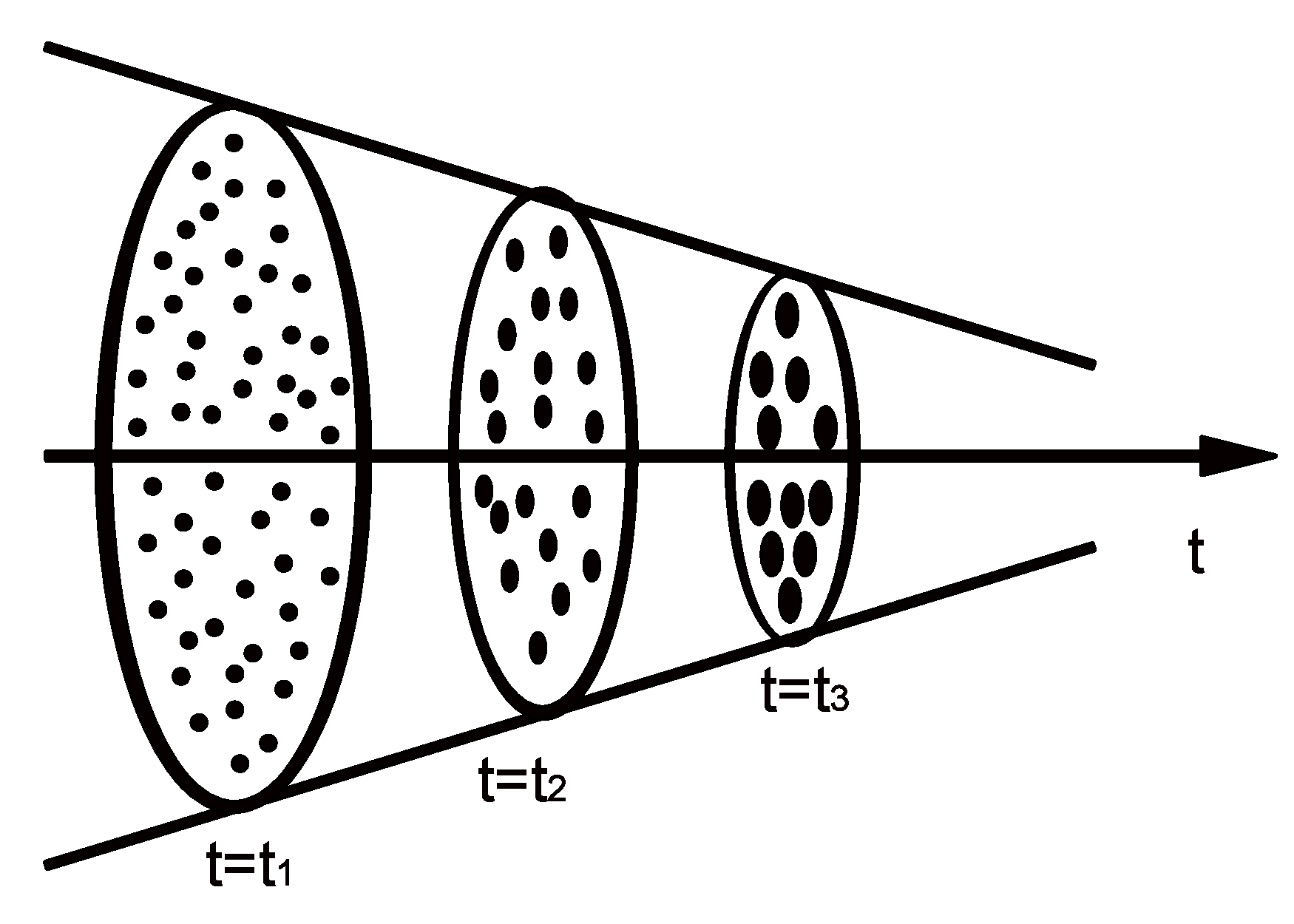}
\caption{A schematic diagram illustrating the idea of self-similarity in the kinetics of 
aggregation with self-replication process. Three circles with progresively smaller
size containg increasingly lesser but larger particle population, which represent snapshots of
 the process at three different times,
are shown similar since the corresponding dimensionless quantities coincide. 
} 
\label{Figure 5}
\end{figure}

The question is: How do we verify Eq. (\ref{eq:scaling_solution}) using the data extracted from numerical simulation? 
First, we need to appreciate the fact that each step of the algorithm does not correspond to one
time unit since time $t\sim 1/((1-p)N)$ in the long-time limit as predicted by Eq. (\ref{eq:m0}).
Second, we collect data for a fixed time $t$ and apreciate the fact that $c_t(x)$ is the histrogram where the height represents
the number of particles within a given range, say of width $\Delta x$, normalized by the width itself so that area under
curve gives the number of particles present in the system at time $t$ regardless of their size. This is exactly
what is shown in Figs. (1) and (2) while the Fig. (2) is shown in the $\log$-linear scale to show that $c_t(x)$
for fixed time decays exponentially.
Now, the solution given by Eq. (\ref{eq:scaling_solution}) implies that distinct data 
points of $c(x,t)$ as a function of $x$ at various different
times can be made to collapse on a single master curve if we plot 
$t^{{{2}\over{(1-p)}}}c(x,t)$ vs $xt^{-{{1+p}\over{(1-p)}}}$ instead. Note that multiplying time $t$ by a constant 
multiplying factor $(1-p)$ has no impact in the resulting plot. 
Indeed, we find that the same data points of all the three distinct curves of
 Fig (2)  merge superbly onto a 
single universal curve, see Fig. (4), which is essentially the scaling function $f(\xi)$. It is clear from Fig. (4) that 
the scaling function $f(\xi)$ decays exponentially and once again this is in perfect aggrement with our analytical solution
given by Eq. (\ref{eq:scaling_function}).

To explain the significance of the data-clollapse better we have drawn in Fig. (5) a schematic
diagram of the process indicating three snapshots at three different times
such that $t_1<t_2<t_3$. The three plots for the distribution function $c(x,t)$ drawn in
Fig (2) may well be considered to represent data extracted from the three snapshots shown in Fig. (5). 
 Now the collapse of the three curves, as shown in Fig. (4),  can only suggest that for a given numerical value of the dimensionless governing quantities
$xt^{-{{1+p}\over{(1-p)}}}$ of the three snapshots, the numerical value of the corresponding dimensionless governed  
quantities $t^{{{2}\over{(1-p)}}}c(x,t)$ of the three snapshots coincide suggesting that the three snapshots 
are simililar. Note that in general two phenomena are called similar if their corresponding dimensionless quantities are identical
which is reminiscent of the fact that two triangles are said to be similar if their respective angles (dimensionless quantities) 
are identical. This is exactly being revealed by the data collapse.

\begin{figure}
\includegraphics[width=8.5cm,height=5.0cm,clip=true]{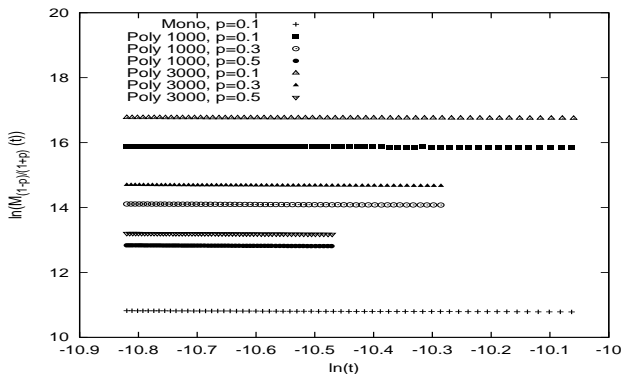}
\caption{$\ln(M_{(\frac{1-p}{1+p})}(t))$ is plotted against $\ln(t)$
for various values of $p$ and various different initial conditions. The horizontal straight lines indicate
that $M_{(\frac{1-p}{1+p})}(t)$ is constant in the scaling regime. In all cases initially 50,000 particles were drawn randomly 
from the size range between $1$ and $n$ where $n=1000,3000$ and denoted as poly n.}
\label{Figure 6}
\end{figure}

We find it instructive to incorporate the scaling solution given by Eq.
(\ref{eq:scaling_solution}) in Eq. (\ref{eq:moment}) to find that
\begin{equation}
\label{eq:scalingMoment}
 M_{j}(t) \sim  ((1-p)t)^{(j-{{1-p}\over{1+p}})z} \Gamma{(j+1)}, 
\end{equation}
as $t \rightarrow \infty$. It is evident from this solution of the $j$th moment that the violation
of the conservation of mass principle is replaced by a non-trivial conservation law
as we find that 
\begin{equation}
\label{eq:conservationlaw}
M_{(\frac{1-p}{1+p})}(t)=\int_0^\infty x^{{{1-p}\over{1+p}}}c(x,t)dx\sim const.
\end{equation}
To verify this using numerical data we label each particle of the system at a given time $t$ 
by the index $i=1,2,3,....,N$ where $N=M_0(t)$ is the total number of particles present in the system at time $t$. 
Then we construct the $q$th moment at time $t$ given by $\sum_ix_i^q$ which is equivalent
to its theoretical counterpart $\int_0^\infty x^qc(x,t)dx$ in the continuum limit. In Fig. (5)
we have shown that the sum of the $q$th power of the sizes of all the existing
particles in the system remain conserved regardless of time $t$ if we choose $q=\frac{1-p}{1+p}$.
Conserved quantities have always attracted physicists as they usually 
point to some underlying symmetry in the
theory or model in which they manifest. Therefore, it is worth
pursuing an understanding of the non-trivial value $\frac{1-p}{1+p}$ for $p>0$ as it
leads to the conserved quantity $M_{(\frac{1-p}{1+p})}(t)$ in the
scaling regime. Such a non-trivial conserved quantity has also been reported in one of our recent works
on condensation-driven aggregation and indicate that it is closely related to the fractal dimension.
It will be interesting if we find similar close connections between fractal dimension and the non-trivial 
conserved quantity.

\section{Fractal Analysis}

\begin{figure}
\includegraphics[width=8.5cm,height=5.0cm,clip=true]{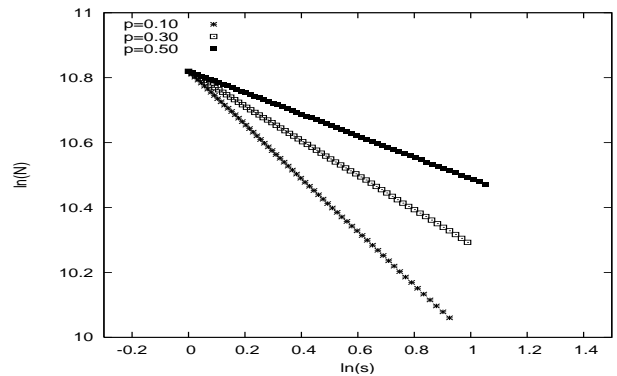}
\caption{Plots of $\ln(N(s))$ against $\ln(s)$ are drawn for three
different values of $p$ for the same initial conditions. The lines
have slopes equal to $-(\frac{1-p}{1+p})$ as predicted by theory. In each case simulation was performed 
till $30,000$ aggregation events while the process started with initially
$N(0)=50,000$ particles of unit size.}
\label{Figure 7}
\end{figure}

\begin{figure}
\includegraphics[width=8.5cm,height=5.0cm,clip=true]{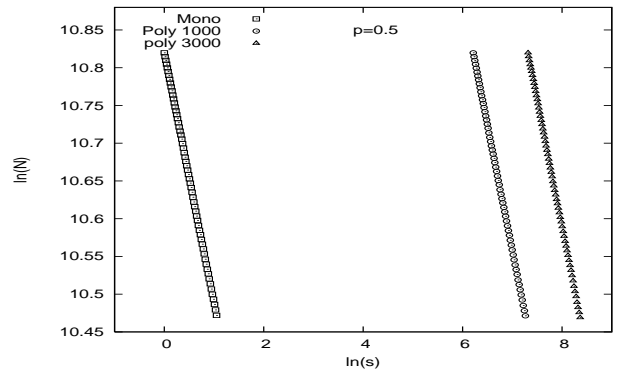}
\caption{The parallel lines resulting from plots of $\ln(N(s))$
against $\ln(s)$ for mono-disperse and poly-disperse initial
conditions confirming that $N(s)\sim s^{-(\frac{1-p}{1+p})}$ is
independent of the initial conditions. In each case simulation started with initially 50,000 particles
were drawn randomly from the size range between 
$1$ and $1000$ for poly 1000, between $1$ and $3000$ for poly 3000 and for monodisperse initial condition all the particles were chosen to be
of unit size.} 
\label{Figure 8}
\end{figure}

In fractal analysis, one usually seeks for a power-law relation between the number $N(\delta)$ 
needed to cover the object under investigation and an yard-stick of size $\delta$ as its exponent $d$ 
gives the geometric dimension of the object. 
It has been found in numerous occasions that besides Euclidean objects that correspond to 
integer exponents $d$ there exist yet another class of objects that correspond to
non-integer exponents $d_f$ of the power-law relation
between $N$ and $\delta$. In the latter case, it has been found that $d_f$ is typically less than the 
dimension of the embedding space and the corresponding object is called fractal \cite{ref.feder}. 
Unlike in Ref. \cite{ref.fractalCDA} here we take a different approach for fractal analysis of the present model. 
Note that the Smoluchowski equation describe aggregation in one dimension so
the idea of collisions of Brownian particles in one dimension is limited to a thought experiment only. 
We then subdivide the system into boxes of size equal to 
that of the respective particles and label them as $i=1,2,...,N$ so that the occupation probability of the $i$th box is
$p_i\propto x_i^{{{1-p}\over{1+p}}}$. 
We then construct the partition function $Z_q$ used typically in the multifractal formalism and it is defined as the $q$th moment of the probability $p_i$ 
\begin{equation}
Z_q=\sum_i^Np_i^q=\sum_i^Nx_i^{{{1-p}\over{1+p}}q}.
\end{equation}
This is in fact the $(1-p)q/(1+p)$th moment of $c(x,t)$ in the continuum limit and hence its solution can be obtained from Eq. (\ref{eq:scalingMoment}) by setting $j=(1-p)q/(1+p)$. Expressing the resulting solution in terms of the mean particle size gives
\begin{equation}
Z_q(s) \sim s^{-\tau(q)},
\end{equation} 
where the mass exponent
\begin{equation}
\tau(q)=(1-q)d_f,
\end{equation}
with $d_f=(1-p)/(1+p)$. 
Note that $\tau(1)=0$ as required by normalization of the probabilities $p_i$s and $\tau(0)=d_f$  is simply the
fractal dimension since we have $Z_0(s)=N(s)$ is the number of yard-stick of size $s$ needed to cover the system and it exhibits power-law 
\begin{equation}
\label{eq:29}
N(s)\sim s^{-d_f}.
\end{equation}
 Notice that the exponent $d_f$ is a non-integer $\forall \ p$ where $0<p<1$ and its value is less than the dimension of the embedding space and hence it is the 
fractal dimension of the resulting system  \cite{ref.feder}. 
To verify our analytical result, we have drawn $\ln (N)$ versus $\ln (s)$ in Fig. (6) from the numerical data collected
for a fixed initial condition but varying only the
$p$ value. On the other hand, in Fig. (7) we have drawn the same plots for a fixed $p$ value but varying only
initial conditions (monodisperse and polydisperse). Both figures show an excellent
power-law fit as predicted by Eq. (\ref{eq:29}) with an exponent exactly equal to $d_f$ regardless of the choice we make
for the initial size distribution of particles in the system.

\section{Discussion and summary}

We have investigated a class of aggregation process
with stochastic self-replication. In the case of mono-disperse initial condition we presented
an  exact analytical solution for the particle size distribution function
$c(x,t)$ and shown that in the limit $t\rightarrow \infty$ it evolves to a dynamic scaling form. 
We then used simple dimensional analysis and the  Backingham $\pi$-theorem to solve the model as it requires
no prior specification of initial condition. To this end, we found that the solution for $c(x,t)$ assumes exactly 
the same dynamic scaling form as the one we found from exact solution for mono-disperse initial condition.
It implies that the dynamic scaling form for 
$c(x,t)$ is universal in the sense that it is independent of initial condition and indeed we have verified it numerically.
Yet another advantage of using the 
Buckingham $\pi$-theorem over the exact solution
is that it provides a processing procedure of verifying the 
dynamic scaling form where the definition of dimensionless quantity is recalled. In particular, we have shown that the distinct
plots of $c(x,t)$ vs $x$ for three different fixed times collapse onto a single universal curve if we plot
the same data in the dimensionless scale namely $t^{{{2}\over{(1-p)}}}c(x,t)$ vs $xt^{-{{1+p}\over{(1-p)}}}$. The collapse of the
distinct curves implies that the systems as it evolves, self-organizes into a self-similar universal state regardless of whether
we choose mono-disperse or poly-disperse initial conditions.     

We have shown crearly that the kinetics of aggregation of particles with self-replication always results in a fractal 
and the value of the fractal dimension $d_f$ is
the same as the index of the conserved moment $\frac{1-p}{1+p}$.  
Such connections between the fractal dimension and the conserved
quantity was first reported by  Ben-Naim and Krapivsky in the context of 
the stochastic Cantor set \cite{ref.fractal}, and later it was found in several other systems 
as well \cite{ref.hassanRogers1,ref.hassanRogers2,ref.hassan1,ref.hassan2,ref.massloss}. 
Recently, Hassan and Hassan have found such connection also in aggregation process 
\cite{ref.fractalCDA}. They have shown that the index of the 
conserved moment is indeed equal to  the fractal
dimension of the resulting system usdergoing condensation-driven aggregation.
We can even apply the idea to the triadic Cantor set, one of the best known textbook example of fractal, 
to check if  the $d_f$th moment, where 
$d_f=\ln 2/\ln 3$, of the remaning intervals is a conserved quantity or not.  It is easy to realize that at the  $n$th generation step
the system consists of $N=2^n$ number of intervals of size $x_i=3^{-n}$. We thus find that the $d_f$th moment of the 
remaining intervals at $n$th generation step is  
\begin{equation}
M_{\ln 2/\ln 3}=\sum_i^{2^n} x_i^{{{1-p}\over{1+p}}}=2^n \Big (3^{-n}\Big )^{{{\ln 2}\over{\ln 3}}}=1,
\end{equation}
independent of $n$.  It once again 
confirms the fact that the fractal dimension $d_f$ is indeed closely connected to the index of the conserved moment.

To further support our fractal analysis, we can use the simple dimensional analysis. According to Eq. (\ref{eq:scaling_solution}) the 
physical dimension of $c(x,t)$ is $[c]=L^{-(1+d_f)}$ since $[s(t)]=L$ and $\theta=1+d_f$. On the other hand, the 
concentration $c(x,t)$ is defined as 
the number of particles per unit volume of embedding space ($V\sim L^d$ where $d=1$) per unit mass ($M$) and hence
$[c]=L^{-1}M^{-1}$  \cite{ref.rajesh}. Now applying the principle of equivalence we obtain
\begin{equation}
\label{eq:masslength}
M(L)\sim L^{d_f}.
\end{equation}
This relation is often regarded as the hallmark for the emergence of fractality.
An object whose mass-length relation satisfies Eq. (\ref{eq:masslength}) with a non-integer exponent is said to be
a fractal in the sense that if the linear dimension of  the object is increased by a factor of $L$ the mass of the
object is not increased by the same factor. That is, the distribution of mass in the object becomes less dense at a
larger length scale. It implies mass exponent $\theta$ is actually the sum of the  dimension of the 
fractal ($d_f$) and that of its embedding space ($d=1$) and it is consistent with the definition of the distribution 
function $c(x,t)$ as well. It is interesting to note that such a simple dimensional analysis can also provide us 
with an answer to the
question: Why is the moment $M_{d_f}=\int_0^\infty x^{d_f}c(x,t)dx$ a conserved quantity?
For an asnwer, we find it conventient to look into the physical dimension of its differential quantity 
$dM_{d_f}=x^{d_f}c(x,t)dx$. Using
the physical dimension $[x]=L$ and $[c(x,t)]=L^{-(1+d_f)}$ in the expression for $dM_{d_f}$, 
we immediately find that it bears no dimension and so does the quantity $M_{d_f}$. Recall that the numerical value
of a dimensionless quantity always remains unchanged upon transition from one unit of measurement
to another within a given class. In the context of the present model it implies that the numerical value of $M_{d_f}$ 
remains the same despite the fact that the system size continues to grow with time.

In summary, besides solving the model analytically, we performed extensive  numerical simulation which fully support
all theoretical findings. Especially, the conditions under which scaling and fractals emerge
are found, the fractal dimension of the system
is given and the relationship between this fractal dimension and a
conserved quantity pointed out. Our findings complement the results found in the condensation-driven 
aggregation indicating
that these results are ubiquitous in the aggregation processes where mass conservation is violated. 
We hope this work will provide
useful insights and theoretical predictions for aggregation processes in
physical, chemical and biological systems with self-replications. 
It would be instructive to analyze our model with other reaction
rates such as sum kernel $K(x,y) = x+y$ and product kernel $K(x,y) = xy$. In the case where
$K(x,y) = xy$, we expect the stochastic self-replication mechanism
to affect the sol-gel phase transition time. We propose to
investigate these issues in subsequent work and hope that the
present work will attract a renewed interest in the subject of
aggregation.

We thank Dr. Naureen Ahsan and  Dr. Arshad Momen for offering critical and useful suggestions after carefully reading the manuscript. 
NI acknowledges support from the Bose Centre for Advanced Study and Research in Natural Sciences.

\end{document}